# Resonant thermoelectric transport in atomic chains with Fano defects

J. Eduardo González[1], Vicenta Sánchez[2], and Chumin Wang[1,]*

[1]Instituto de Investigaciones en Materiales, Universidad Nacional Autónoma de México, 04510 Mexico City, Mexico
[2]Departamento de Física, Facultad de Ciencias, Universidad Nacional Autónoma de México, 04510 Mexico City, Mexico
*Corresponding author: e-mail: chumin@unam.mx, Phone: +52-55-56224634, Fax: +52-55-56161251

**Abstract**

Atomic clusters attached to a low-dimensional system, called Fano defects, produce rich wave interferences. In this work, we analytically found an enhanced thermoelectric figure-of-merit (*ZT*) in periodic atomic chains with Fano defects, compared to those without such defects. We further study self-assembled DNA-like systems with periodic and quasiperiodically placed Fano defects by using a real-space renormalization method developed for the Kubo-Greenwood formula, in which tight-binding and Born models are respectively used for the electric and lattice thermal conductivities. The results reveal that the quasiperiodicity could be another *ZT*-improving factor, whose long-range disorder inhibits low-frequency acoustic phonons insensitive to local defects.

**I. Introduction**.

Defects and impurities in semiconductors have decisive effects on their transport and optical properties, as demonstrated in current microelectronic devices. Recently, branched nanowires have captured considerable attention due to their unique physical properties for applications like photovoltaics and catalysis [1]. On the theoretical side, such branches can be modeled as Fano defects that consist of atomic chains joined to a low-dimensional system [2]. In the last years, the existence of null and ballistic conduction states in periodic chains with a single Fano defect is found [3]. Moreover, a novel ballistic conduction state in two-dimensional belts with a non-periodic arrangement of atoms along the Fano plane defect has been analytically proved [4]. In addition, enhancements to the ballistic alternating current (AC) conductivity are also reported when quasiperiodically placed Fano defects are introduced to a periodic chain or nanowire [5].

These peculiar transport properties of Fano systems could be used for clean energy conversion through thermoelectric devices, including deoxyribonucleic acid (DNA) based self-assembled systems [6]. The efficiency of thermoelectric devices is generally determined through the dimensionless thermoelectric figure-of-merit $ZT \equiv \sigma S^2 T/(\kappa_{el}+\kappa_{ph})$, where $\sigma$, $S$, $T$, $\kappa_{el}$ and $\kappa_{ph}$ are respectively electrical conductivity, Seebeck coefficient, temperature, thermal conductivities by electrons and by phonons [7]. These thermoelectric quantities can be determined by using the Boltzmann formalism [8]. Nevertheless, the inherent correlation between them, as exposed by the Wiedemann-Franz law for metals, makes it difficult to improve the value of *ZT*. A recent study of poly(G)-poly(C) DNA chains without including $\kappa_{ph}$ predicts a huge *ZT* induced by the sugar-phosphate backbones [9]. However, the presence of $\kappa_{ph}$ in *ZT* could be decisive specially for $\kappa_{el} \approx 0$ when the chemical potential ($\mu$) is located outside the electronic conducting band. In this article, we report novel analytical results of *ZT* obtained from the Kubo-Greenwood formula applied to both electronic and phononic transports in infinite periodic chains with and without Fano impurities. Moreover, numerical calculations of *ZT* in periodic and quasiperiodic DNA systems as well as their comparison with previous works are presented.

## II. The Boltzmann-Kubo model

Let us consider a linear chain of *N* atoms along the *x*-direction with $N_F$ single-atom Fano defects perpendicularly attached to each of them. The single-band tight-binding Hamiltonian of this chain can be written as

$$\hat{H} = \sum_{j=1}^{N} \left( \varepsilon_j |j\rangle\langle j| + t_{j,j+1} |j\rangle\langle j+1| + t_{j,j-1} |j\rangle\langle j-1| + \sum_{s=1}^{N_F} \left( \varepsilon_s |s\rangle\langle s| + t_{j,s} |j\rangle\langle s| + t_{j,s} |s\rangle\langle j| \right) \right), \quad (1)$$

where $\varepsilon_j$ is the self-energy of *j*-th atom in the chain, $\varepsilon_s$ denotes the site-energy of its *s*-th Fano defect, $t_{j,j+1}$ represents the hopping integral between the *j*-th atom and its nearest-neighbor atom *j*+1, and $t_{j,s}$ specifies the hopping integral between the *j*-th atom and its *s*-th Fano impurity. Hence, the linear momentum along the *x* axis is

$$\hat{p}_x = \frac{im}{\hbar}[\hat{H}, \hat{x}] = \frac{ima}{\hbar} \sum_j \left( t_{j,j+1} |j\rangle\langle j+1| - t_{j,j-1} |j\rangle\langle j-1| \right), \quad (2)$$

where *m* is the mass of electrons, *a* is the interatomic distance, $\hat{x} = \sum_j x_j |j\rangle\langle j|$ and $x_j = ja$ is the *x*-axis coordinate of *j*-th atom.

Thermoelectric quantities within the Boltzmann formalism are given by [8],

$$\sigma(\mu, T) = e^2 L_0(\mu, T), \quad (3)$$

$$S(\mu, T) = -\frac{L_1(\mu, T)}{|e|T L_0(\mu, T)}, \quad (4)$$

$$\kappa_{el}(\mu, T) = \frac{1}{T} \left( L_2(\mu, T) - \frac{L_1^2(\mu, T)}{L_0(\mu, T)} \right), \quad (5)$$

and

$$ZT(\mu, T) = \frac{\sigma(\mu, T) S^2(\mu, T) T}{\kappa_{el}(\mu, T) + \kappa_{ph}(T)} = \frac{L_1^2(\mu, T) / L_0(\mu, T)}{L_2(\mu, T) - L_1^2(\mu, T) / L_0(\mu, T) + T\kappa_{ph}(T)}, \quad (6)$$

where *e* is the electric charge of electrons and

$$L_n(\mu, T) = \frac{-2\hbar}{\pi m^2 \Omega} \int_{-\infty}^{\infty} dE (E-\mu)^n \frac{\partial f}{\partial E} Tr\left\{ \hat{p}_x \, \text{Im} G^+(E) \hat{p}_x \, \text{Im} G^+(E) \right\} \quad (7)$$

is calculated by means of the Kubo-Greenwood formula [10], in which $\Omega$ is the system volume, $f(E) = \{1 + \exp[(E-\mu)/(k_B T)]\}^{-1}$ is the Fermi-Dirac distribution with chemical potential $\mu$ and temperature $T$, $G^+(E) = \lim_{\eta \to 0^+} G(E+i\eta)$ is the retarded electronic Green's function determined by the Dyson's equation $(z-\hat{H})G(z) = 1$ with $z = E + i\eta$ [11]. In Eq. (6), the thermal conductivity by phonons $\kappa_{ph}(T)$ can be calculated by using the Kubo-Greenwood formula for phonons given by [12]

$$\kappa_{ph}(T) = \frac{-2\hbar^2}{\pi \Omega k_B T^2} \sum_{l=1}^{3} \int_0^\infty \frac{\omega^2 e^{\hbar\omega/k_B T} d\omega}{(e^{\hbar\omega/k_B T}-1)^2} Tr\{\mathbf{A}_x \mathrm{Im} G_{ph}^+(\omega^2) \mathbf{A}_x \mathrm{Im} G_{ph}^+(\omega^2)\}_l, \tag{8}$$

where the summation of $l$ is over a longitudinal (L) and two transversal vibrational modes (T), $G_{ph}^+(\omega^2)$ is the retarded phononic Green's function determined by the Dyson's equation $(\omega^2 \mathbf{I} - \mathbf{\Phi})G_{ph}(\omega^2) = \mathbf{I}$ with $\mathbf{I}$ the identity matrix and the element $(\mu, \nu)$ of matrix $\mathbf{A}_x$ is given by

$$[\mathbf{A}_x]_{\mu,\nu}(j,k) = \frac{1}{2}(\mathbf{R}_j - \mathbf{R}_k)_x \Phi_{\mu,\nu}(j,k) = \frac{(\mathbf{R}_j - \mathbf{R}_k)_x}{2\sqrt{M_j M_k}} \frac{\partial^2 V(j,k)}{\partial u_\mu(j) \partial u_\nu(k)}, \tag{9}$$

in which $\mathbf{\Phi}$ is the dynamic matrix, $\mathbf{R}_j$ ($M_j$) and $\mathbf{R}_k$ ($M_k$) are equilibrium positions (mass) of atoms $j$ and $k$, respectively. The interaction potential $V(j,k)$ between nearest-neighbor atoms $j$ and $k$ within the Born model is given by [13]

$$V(j,k) = \frac{\alpha_{j,k} - \beta_{j,k}}{2} |[\mathbf{u}(j) - \mathbf{u}(k)] \cdot \hat{\mathbf{n}}_{j,k}|^2 + \frac{\beta_{j,k}}{2} |\mathbf{u}(j) - \mathbf{u}(k)|^2, \tag{10}$$

where $\mathbf{u}(j)$ is the displacement of $j$-th atom with respect to its equilibrium position, $\alpha_{j,k}$ and $\beta_{j,k}$ are respectively the central and non-central restoring force constants between atoms $j$ and $k$, whose bond direction is indicated by the unitary vector $\hat{\mathbf{n}}_{j,k}$. In Eq. (10), index $k$ runs over all nearest neighbors of $j$-th atom, including the Fano defect atoms.

### III. Analytical study

For a periodic linear chain without Fano defects of $N$ atoms with mass $M$, null self-energies, constant hopping integral ($t$), uniform central ($\alpha$) and non-central ($\beta$) restoring force constants, connected to two semi-infinite periodic leads with the same parameters at its ends, as shown in the inset of FIG. 1(c), the traces of Eqs. (7) and (8) can be analytically evaluated as [14]

$$Tr\{\hat{p}_x \mathrm{Im} G^+(E) \hat{p}_x \mathrm{Im} G^+(E)\} = \frac{(N-1)^2 a^2 m^2}{2\hbar^2} \Theta(2|t|-|E|), \tag{11}$$

and

$$Tr\{\mathbf{A}_x \mathrm{Im} G_{ph}^+(\omega^2) \mathbf{A}_x \mathrm{Im} G_{ph}^+(\omega^2)\}_l = -\frac{(N-1)^2 a^2}{8} \Theta(4\omega_l^2 - \omega^2), \tag{12}$$

where $\Theta(x) = \begin{cases} 0, & \text{if } x < 0 \\ 1, & \text{if } x \geq 0 \end{cases}$ is the step function and $\omega_l = \begin{cases} \sqrt{\alpha/M}, & \text{if } l = \text{L} \\ \sqrt{\beta/M}, & \text{if } l = \text{T} \end{cases}$.

In consequence, Eq. (7) gives the rise to

$$L_0^{1D}(\mu, T) = A(-2|t|-\mu) - A(2|t|-\mu), \tag{13}$$

$$L_1^{1D}(\mu,T) = B(-2|t|-\mu) - B(2|t|-\mu),  \tag{14}$$

$$L_2^{1D}(\mu,T) = \begin{cases} C(-2|t|-\mu) - C(2|t|-\mu), & \text{if } \mu < -2|t| \\ \dfrac{\pi \Omega k_B^2 T^2}{3\hbar} - C(2|t|-\mu) - C(\mu+2|t|), & \text{if } |\mu| < 2|t| \\ C(\mu-2|t|) - C(\mu+2|t|), & \text{if } \mu > 2|t| \end{cases}, \tag{15}$$

where $\Omega = (N-1)a$ is the linear chain length,

$$A(x) = \frac{\Omega}{\pi\hbar} \frac{1}{e^{x/k_BT}+1}, \tag{16}$$

$$B(x) = \frac{\Omega}{\pi\hbar}\left[\frac{x}{e^{x/k_BT}+1} + k_BT \ln(e^{-x/k_BT}+1)\right], \tag{17}$$

and

$$C(x) = \frac{\Omega}{\pi\hbar}\left[\frac{x^2}{e^{x/k_BT}+1} + 2xk_BT\ln(e^{-x/k_BT}+1) - 2(k_BT)^2 \lim_{Q\to\infty}\sum_{s=1}^{Q}\frac{e^{-sx/k_BT}}{(-1)^s s^2}\right]. \tag{18}$$

Moreover, equations (8) and (12) lead to

$$\kappa_{ph}^{1D}(T) = \frac{\Omega k_B}{\pi\hbar} \sum_{l=1}^{3}\left[\frac{\pi^2 k_B T}{12} + D(2\omega_l)\right], \tag{19}$$

where

$$D(\omega) = \frac{\hbar^2\omega^2 e^{-\hbar\omega/k_BT}}{4k_BT(e^{-\hbar\omega/k_BT}-1)} + \frac{\hbar\omega}{2}\ln(1-e^{-\hbar\omega/k_BT}) - \frac{k_BT}{2}\lim_{P\to\infty}\sum_{s=1}^{P}\frac{e^{-s\hbar\omega/k_BT}}{s^2}. \tag{20}$$

From Eqs. (13-15) and (19) we can define dimensionless thermoelectric quantities by underlined symbols given by

$$\underline{L}_n^{1D} = L_n^{1D} \frac{\pi\hbar}{\Omega|t|^n} \tag{21}$$

and

$$\underline{\kappa}_{ph}^{1D} = \kappa_{ph}^{1D} \frac{\pi\hbar}{\Omega k_B |t|}, \tag{22}$$

for $n = 0, 1$ and $2$. In FIGS. 1(a-d), we compare the analytical solutions (open symbols) of (a) $L_2^{1D}$, (b) $\kappa_{ph}^{1D}$, (c) $ZT_{el}$ and (d) $ZT$ with their respective numerical results (blue lines) as functions of the temperature ($T$) for the periodic chain shown in the inset of FIG. 1(c) with a chemical potential located at $\mu = -2.08|t|$, $|t| = 1$ eV, a site mass of $5.12\times 10^{-25}$ kg, central and non-central restoring force constants of $\alpha = 31.1$ N/m and $\beta = 5.36$ N/m, as occurred in a DNA-like structure discussed in the next section. The numerical calculations were carried out in macroscopic length systems by using an exact renormalization procedure developed for periodic and quasiperiodic lattices [15]. In FIG. 1(c), $ZT_{el}$ denotes the dimensionless thermoelectric figure of merit by electrons, neglecting the phonon participation as in Ref. [9], given by

$$ZT_{el}(\mu,T) = \frac{\sigma(\mu,T)S^2(\mu,T)T}{\kappa_{el}(\mu,T)} = \frac{L_1^2(\mu,T)/L_0(\mu,T)}{L_2(\mu,T) - L_1^2(\mu,T)/L_0(\mu,T)}. \tag{23}$$

The analytical solutions of $L_2^{1D}$ and $ZT_{el}$, red open circles in FIGS. 1(a) and 1(c), were obtained from Eqs. (13-23), whose summations were performed until $Q=1$, while in FIG. 1(b) the analytical solutions of $\kappa_{ph}^{1D}$ were gotten from Eq. (19) whose summation was truncated at $P=1$ (blue open triangles), $P=2$ (green open rhombuses) and $P=20$ (red open circles). The corresponding analytical solutions of $ZT$ are showed in FIG. 1(d).

Notice the remarkable accordance between the numeric and analytic results, as well as the number of summation terms required in the analytical solutions, *i.e.*, $Q=1$ for $L_2^{1D}$ while $P=20$ for $\kappa_{ph}^{1D}$. This truncation difference between electric and thermal contributions is originated from electron and phonon characteristic energies respectively given by $|t|$ and $\hbar\omega_l$, where $|t|\gg\hbar\omega_l$ leading to a reduced summation-term requirement for the electron case. Moreover, observe the divergence of $ZT_{el}$ in FIG. 1(c) when the temperature goes to zero, following

$$ZT_{el}(\mu,T) \approx \frac{(-\mu-2|t|+k_BT)^2}{(k_BT)^2}, \quad \text{for } \mu<-2|t| \text{ and } T\to 0, \tag{24}$$

which can be obtained by taking $\ln[e^{(\mu+2|t|)/k_BT}+1]\approx e^{(\mu+2|t|)/k_BT}$, $1/[e^{(-\mu-2|t|)/k_BT}+1]\approx e^{(\mu+2|t|)/k_BT}$, $Q=1$ in Eq. (18), and neglecting $A(2|t|-\mu)$, $B(2|t|-\mu)$ and $C(2|t|-\mu)$ in Eqs. (13-15), since $\mu=-2.08|t|$ is close to the lower band edge ($-2|t|$) and far from the upper one ($2|t|$). In contrast, $ZT$ goes to zero when $T$ decreases in FIG. 1(d), due to the presence of thermal conductivity by phonons in Eq. (6) which can be expressed as $\kappa_{ph}^{1D}\approx\Omega\pi k_B^2 T/(4\hbar)$ at very low temperatures [14]. Then, Eq. (6) leads to

$$ZT(\mu,T) \approx \frac{ZT_{el}(\mu,T)}{1+\exp[(-\mu-2|t|)/k_BT]\pi^2/4}, \quad \text{for } \mu<-2|t| \text{ and } T\to 0. \tag{25}$$

Now, we consider an infinite periodic chain with two single-atom Fano defects attached to each atom of the main chain, which is connected to two semi-infinite periodic leads with the same Fano impurities, as shown in the inset of FIG. 1(e). All these atoms have a null self-energy and are linked with their nearest neighbors through a hopping integral *t*, central (*α*) and non-central (*β*) restoring force constants. The coordinates of these two single-atom Fano defects can be renormalized producing an effective self-energy ($\varepsilon_{imp}=2t^2/E$) at atoms on the main chain [16] and then, the original problem is reduced to an equivalent one-dimensional (1D) linear chain. This Fano-defect induced self-energy leads to new electronic band edges located at $E_1=-|t|(\sqrt{3}+1)$, $E_2=-|t|(\sqrt{3}-1)$, $E_3=|t|(\sqrt{3}-1)$ and $E_4=|t|(\sqrt{3}+1)$, being solutions of $E-2t^2/E=\pm 2|t|$. Hence, the trace of Eq. (7) can be written as

$$Tr\{\hat{p}_x \text{Im} G^+(E)\hat{p}_x \text{Im} G^+(E)\} = \frac{(N-1)^2 a^2 m^2}{2\hbar^2}\left[\Theta\left(|t|-\left|E+\sqrt{3}|t|\right|\right)+\Theta\left(|t|-\left|E-\sqrt{3}|t|\right|\right)\right]. \tag{26}$$

There are three different vibrational modes in this periodic chain with Fano defects, and they are longitudinal (L), transversal along the Fano impurities (T1), and transversal perpendicular to the Fano impurities (T2). The trace of Eq. (8) is given by

$$Tr\{\mathbf{A}_x \text{Im} G^+_{ph}(\omega^2)\mathbf{A}_x \text{Im} G^+_{ph}(\omega^2)\}_l = -\frac{(N-1)^2 a^2}{8}\{\Theta[(\omega_1^l)^2-\omega^2]+\Theta[(\omega_3^l)^2-\omega^2]\Theta[\omega^2-(\omega_2^l)^2]\}, \quad (27)$$

where the phononic band edges ($\omega_1^l$, $\omega_2^l$ and $\omega_3^l$) are presented in Table 1 for the three mentioned vibrational modes.

Evaluating Eqs. (7-8) for this periodic chain with Fano defects we obtain

$$L_0^F(\mu,T) = A(E_1-\mu) - A(E_2-\mu) + A(E_3-\mu) - A(E_4-\mu), \quad (28)$$

$$L_1^F(\mu,T) = B(E_1-\mu) - B(E_2-\mu) + B(E_3-\mu) - B(E_4-\mu), \quad (29)$$

$$L_2^F(\mu,T) = \begin{cases} C(E_1-\mu) - C(E_2-\mu) + C(E_3-\mu) - C(E_4-\mu), & \text{if } \mu < E_1 \\ \dfrac{\pi\Omega(k_B T)^2}{3\hbar} - C(E_2-\mu) - C(\mu-E_1) + C(E_3-\mu) - C(E_4-\mu), & \text{if } E_1 \leq |\mu| < E_2 \\ C(\mu-E_2) - C(\mu-E_1) + C(E_3-\mu) - C(E_4-\mu), & \text{if } E_2 \leq |\mu| < E_3 , \\ \dfrac{\pi\Omega(k_B T)^2}{3\hbar} - C(\mu-E_3) - C(E_4-\mu) + C(\mu-E_2) - C(\mu-E_1), & \text{if } E_3 \leq |\mu| < E_4 \\ C(\mu-E_2) - C(\mu-E_1) + C(\mu-E_4) - C(\mu-E_3), & \text{if } \mu \geq E_4 \end{cases} \quad (30)$$

and

$$\kappa_{ph}^F(T) = \frac{\Omega k_B}{\pi\hbar}\sum_{l=1}^{3}\left[\frac{\pi^2 k_B T}{12} + D(\omega_1^l) - D(\omega_2^l) + D(\omega_3^l)\right]. \quad (31)$$

In FIG. 1, the dimensionless (e) $\underline{L}_0$ and (f) $ZT$ as functions of the chemical potential μ obtained from Eqs. (28-31) are comparatively presented for infinite periodic chains with (circles and up triangles) and without (squares and down triangles) Fano impurities at $k_B T = 0.01|t|$ (circles and squares) and $k_B T = 0.03|t|$ (up and down triangles). Observe in FIG. 1(e) the appearance of a central energy gap of $2(\sqrt{3}-1)|t|$ around $\mu = 0$ induced by the Fano defects and two shifted bands conserving their original band width of $4|t|$. Note also the height and full-width-at-half-maximum (FWHM) of $ZT$ peaks around each band edge increase with the temperature, while their location shifts away from the band edge. Moreover, the maximum $ZT$ grows with the presence of Fano defects.

## IV. Application to DNA-like systems

The model described in section II can be applied to address the thermoelectric properties of DNA molecules, such as a poly(G)-poly(C) DNA chain schematically illustrated in FIG. 2(f). These molecules can be self-assembled to create long double chains [17]. One of the most used DNA semi-empirical approaches for electrons is the so-called fishbone model [18], in which each base pair, guanine-cytosine (GC) or adenine-thymine (AT), is considered as a single site with self-energy $\varepsilon_0$. Such model also considers hopping integrals $t$ of π-electrons between nearest-neighbor base pairs as well as between the nucleobase and sugar-phosphate backbone represented as Fano defects, whose self-energies depend on the environmental condition [19] and are denoted by $\varepsilon_+$ and $\varepsilon_-$ in FIG. 2(g). These backbones are crucial for the electronic transport in DNA double helix, since they can induce a metal-insulator transition [20, 21]. In fact, the semiconducting behavior of DNA molecules can be used for thermoelectric applications [22]. For example, a notable growth of the Seebeck coefficient was observed when a couple of AT base pairs is introduced into a DNA molecule with GC sequence [23]. Recently, a giant thermoelectric figure of merit by electrons ($ZT_{el}$) induced by the sugar-phosphate backbones in a finite two-dimensional (2D) poly(G)-poly(C) DNA chain was predicted [9].

In this section, we present a full semi-empirical calculation of thermoelectric properties of DNA molecules that includes the phonon participation through a 2D coarse grain model of two sites per nucleotide, similar to that of Ref. [24]. Our two-site coarse-grain model consists of one site representing the nucleobase that could be guanine, cytosine, adenine or thymine, and the other symbolizing the sugar-phosphate molecule, as presented in FIG. 2(h). The interaction between these two sites are modeled by central ($\alpha'$) and non-central ($\beta'$) restoring force constants, while those between nearest-neighbor sugar-phosphate molecules are characterized by $\alpha = 31.1\,\text{N/m}$ and $\beta = 5.36\,\text{N/m}$. These last two constants were calculated from DNA longitudinal ($v_\text{L} = 2650\,\text{m/s}$) and transversal ($v_\text{T} = 1100\,\text{m/s}$) sound velocities estimated by the inelastic X-ray scattering [25], through the relations $v_\text{L} = a\sqrt{\alpha/(M_{SP}+M_B)}$ and $v_\text{T} = a\sqrt{\beta/(M_{SP}+M_B)}$ where $a = 3.4\times10^{-10}\,\text{m}$ is the distance between nearest-neighbor base pairs, $M_{SP} \approx 2.96\times10^{-25}\,\text{kg}$ is the mass of sugar-phosphate molecule and $M_B = (M_G+M_C+M_A+M_T)/4 \approx 2.19\times10^{-25}\,\text{kg}$ is the average mass of four-type nucleobases, since they are almost randomly presented in a DNA double helix. In this coarse-grain model, the interactions between complementary nucleobases and between stacked bases via hydrogen bonds are omitted, since the corresponding force constants are typically smaller than 0.1 N/m [26]. Hence, the vibrational modes of poly(G)-poly(C) DNA chains can be described through two independent chains with guanine and cytosine Fano defects, as shown in FIG. 2(h).

In FIGS. 2(a-e), we present the results denoted by red spheres of (a) 1D thermal conductance by electrons and by phonons (magenta dashed line) defined as $K(\mu,T) = \kappa(\mu,T)\Omega_\perp/\Omega_\parallel$, (b) 1D electrical conductance given by $G(\mu,T) = \sigma(\mu,T)\Omega_\perp/\Omega_\parallel$, (c) Seebeck coefficient ($S$), (d) electronic ($ZT_{el}$) and (e) full ($ZT$) dimensionless thermoelectric figures of merit as functions of the base pair self-energy ($\varepsilon_0$) with $\mu = 0$, $\varepsilon_\pm = \varepsilon_0 \pm 0.1|t|$, $\alpha' = \alpha$, $\beta' = \beta$ and at $k_B T = 0.01|t|$ for a poly(G)-poly(C) DNA chain of five GC base pairs connected to two semi-infinite atomic leads with $\tilde{\varepsilon} = 0$, $\tilde{t} = 2t$ and $M_{\tilde{SP}} = M_{SP}$, as illustrated in FIG. 2(g). In FIGS. 2(a-d), the results of Ref. [9] are plotted as blue lines, in order to compare them with our calculations.

Note the good coincidence of $ZT_{el}$ from Ref. [9] and our calculations, whose difference observed in FIG. 2(b) is due to the dissimilarity of modeling the leads. In addition, we present a full calculation of *ZT* including the phonon contribution to Eq. (6), which generates a realistic small value of *ZT* shown in FIG. 2(e). The initial value of $ZT_{el} \approx 1200$ is originated from an almost null denominator in Eq. (23) and the inclusion of a finite thermal conductivity by phonons [see the dashed line in FIG. 2(a)] into Eq. (6) rectifies this effect and produces a final $ZT \approx 0.75$. It is worth mentioning that in FIG. 2 we have considered $\alpha' = \alpha$ and $\beta' = \beta$, as assumed in Ref. [27], and the effects of these parameters on the *ZT* will be analyzed in detail.

In FIG. 3, we present the *ZT* variation as a function of vibrational parameters $\alpha$ and $\beta$ for $\alpha'/\alpha = \beta'/\beta = 1$ and $\alpha'/\alpha = \beta'/\beta = 0$ analyzed in the same poly(G)-poly(C) DNA chain of FIG. 2 with the nucleobase self-energy $\varepsilon_0 = -0.027|t|$ corresponding to the maximal *ZT* in FIG. 2(e). Notice that *ZT* grows with the decrease of $\alpha$ and/or $\beta$, since it reduces the phononic band width and then its participation in the thermal conductivity. Also, observe the diminution of *ZT* when $\alpha' = \beta' = 0$, which corresponds to the case without phonon Fano defects, whose presence induces a phononic central band gap, as occurred in the electronic case shown in FIG. 1(e). Such band gap diminishes the thermal conductivity by phonons and then yields a larger *ZT*. For the limiting case of $\alpha = \beta = \alpha' = \beta' = 0$, *ZT* should recover the value of $ZT_{el}$.

**IV. Quasiperiodic poly(G)-poly(C) double chains**

In this section, we study quasiperiodically ordered poly(G)-poly(C) double chains following the Fibonacci sequence given by ABAAB… [14], as shown in FIG. 4(a) being A=(GC) and B=(CG) base pairs, and we analyse the effects of quasiperiodicity on the thermoelectric figure of merit *ZT*. The electronic transport in this quasiperiodic double chain is modelled by the fishbone model, FIG. 4(b), while its lattice thermal conductivity is described by means of the previously discussed two-site coarse-grain model, schematically presented in FIG. 4(c). Notice that the fishbone model is insensitive to the (GC) base-pair ordering, *i.e.*, the structures of FIGS. 2(g) and 4(b) are described by the same electronic Hamiltonian of Eq. (1). In contrast, the quasiperiodic order of (GC) base pairs modifies the dynamic matrix of phonons.

FIG. 4(d) shows the thermoelectric figure of merit (*ZT*) versus the nucleobase self-energy ($\varepsilon_0$) for a quasiperiodic poly(G)-poly(C) chain of 1346269 base pairs (red open circles), corresponding to the generation 30 of the Fibonacci sequence, with $\mu = 0$, $\varepsilon_\pm = \varepsilon_0 \pm 0.2|t|$, $\alpha' = \alpha$, $\beta' = \beta$, $k_B T = 0.02|t|$ and the same periodic leads as in FIG. 2. Such results are compared to those obtained from the periodic poly(G)-poly(C) double chain of the same length (blue open squares) schematically shown in FIG. 2(f). Observe in FIG. 4(d) the enhancement of *ZT* when the quasiperiodicity is introduced. Notice also the growth of *ZT* from 0.75 to 2.5 for the periodic case archived at $\varepsilon_0 = 0.0718|t|$ in a longer system at a higher temperature. The effects of system length and temperature on *ZT* are analysed in FIG. 4(e), where a growth of *ZT* with the system length (*N*) and temperature (*T*) is noted in both periodic (blue solid squares) and quasiperiodic (red solid circles)

poly(G)-poly(C) double chains studied in FIG. 4(d). However, there is a qualitative difference between the periodic and quasiperiodic cases in the growth of *ZT* with *N*, *i.e*., a continue progress for the Fibonacci systems versus an almost constant behaviour for the periodic ones. Such contrast is derived from the suppression of low-frequency phonon transport by the long-range quasiperiodic structural disorder.

**V. Conclusions**

The analytical solutions for infinite periodic chains with and without Fano impurities reported in this article let us visualize the contribution of each component of *ZT* and the intrinsic correlations between $L_0$, $L_1$ and $L_2$, as established in the Wiedemann-Franz law for metals. For example, simple analytical expressions of $ZT_{el}$ and *ZT*, respectively given by Eqs. (24) and (25) for low temperature regime, summarize such intrinsic correlations and are obtained from analytical cancelations of distinct terms of *ZT*. Frequently, these cancelations are difficult to numerically carry out, since it involves the summation of quantities that differ by more than forty orders of magnitude in the numerator and/or denominator of *ZT*. Hence, analytical solutions of the limiting case could serve as a guide for the analysis of numerical results, although they are valid only for periodic systems.

These analytical solutions were confirmed by numerical calculations and both predict a general improved *ZT* of periodic chains with Fano impurities, due to the reduction of thermal conductivity by phonons caused by the appearance of a central band gap in the phononic density of states $DOS_{ph}(\omega^2)$. This effect is even enhanced if a quasiperiodic disorder is introduced, since it highly diminishes the thermal conduction of long wavelength acoustic phonons, which are responsible of the phonon transport at low temperature and not easy to be suppressed since they do not feel local defects neither impurities [28].

In addition, we apply this analysis of Fano impurities to a poly(G)-poly(C) DNA system by means of simple coarse-grained models for both electron and phonon descriptions. The results of *ZT* reveal a crucial participation of vibrational contributions, decreasing from $ZT_{el} \approx 1200$ to $ZT \approx 0.75$, where the central ($\alpha$) and non-central ($\beta$) restoring force parameters along the double strands were obtained from the DNA sound velocity measurements while $\alpha'$ and $\beta'$ have a limited influence to the resulting *ZT* as shown in FIG. 3. Finally, a non-perturbative study of macroscopic-length poly(G)-poly(C) DNA chains with quasiperiodic Fibonacci sequence by alternating (GC) and (CG) base-pair order reveals an additional improvement of *ZT* with respect the periodic case by more than 30% for a macroscopic-length system.

**Acknowledgements** This work has been partially supported by CONACyT-252943, UNAM-PAPIIT-IN114916 and UNAM-PAPIIT-IN116317. Computations were performed at Miztli of DGTIC-UNAM.

Table 1. Phononic band edges for a periodic chain with Fano defects

| | $l = \text{L}$ | $l = \text{T1}$ | $l = \text{T2}$ |
|---|---|---|---|
| $\omega_1^l$ | $\sqrt{[4\alpha+3\beta-\sqrt{(4\alpha+3\beta)^2-16\alpha\beta}]/2M}$ | $\sqrt{[4\beta+3\alpha-\sqrt{(4\beta+3\alpha)^2-16\alpha\beta}]/2M}$ | $\sqrt{(7-\sqrt{33})\beta/2M}$ |
| $\omega_2^l$ | $\sqrt{3\beta/M}$ | $\sqrt{3\alpha/M}$ | $\sqrt{3\beta/M}$ |
| $\omega_3^l$ | $\sqrt{[4\alpha+3\beta+\sqrt{(4\alpha+3\beta)^2-16\alpha\beta}]/2M}$ | $\sqrt{[4\beta+3\alpha+\sqrt{(4\beta+3\alpha)^2-16\alpha\beta}]/2M}$ | $\sqrt{(7+\sqrt{33})\beta/2M}$ |

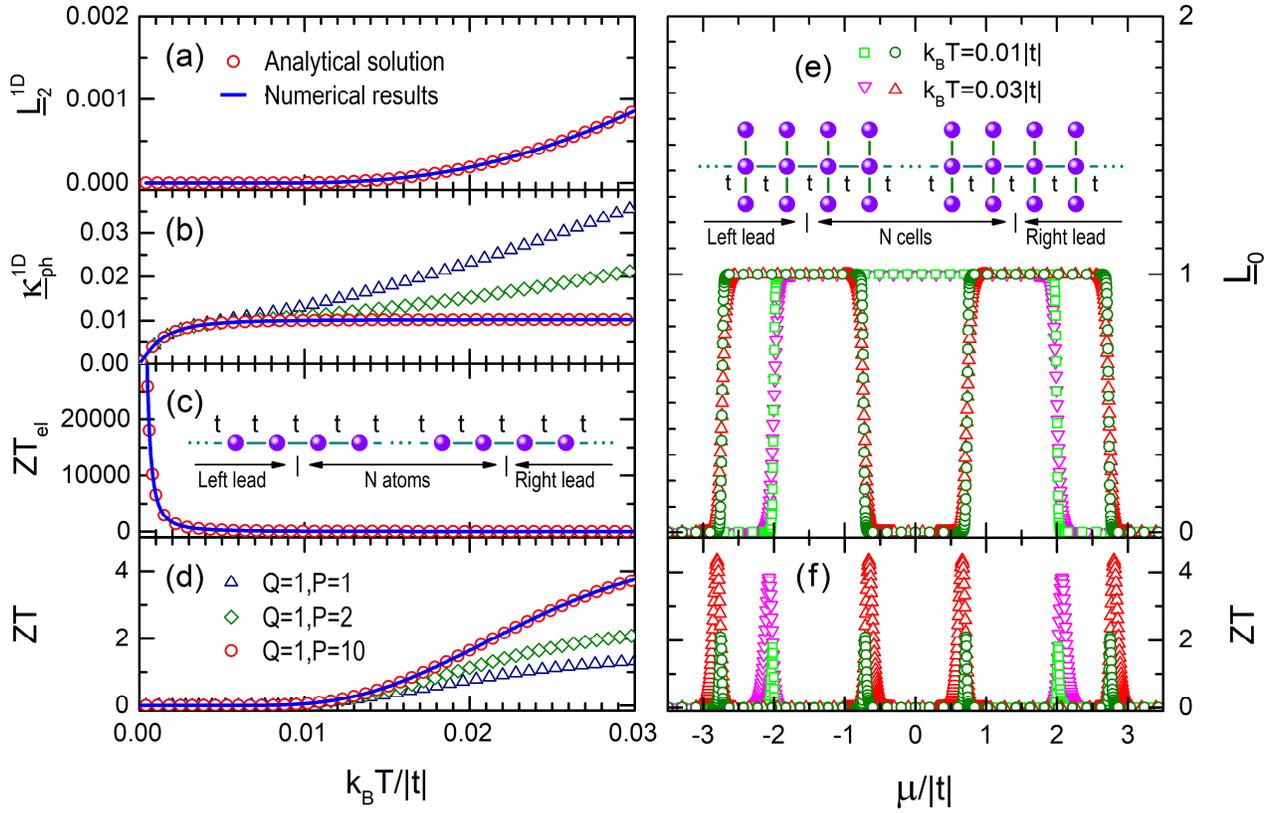

**FIG. 1**. Analytical and numerical results of dimensionless thermoelectric quantities (a) $\underline{L}_2^{1D}$, (b) $\underline{\kappa}_{ph}^{1D}$, (c) $ZT_{el}$ and (d) $ZT$, as functions of the temperature ($T$) for a periodic linear chain connected to two semi-infinite periodic leads for $\mu = -2.08|t|$, where the analytical ones were obtained from Eqs. (6-23) truncating their summations until $Q = 1$ and $P = 1$ (blue open triangles), $P = 2$ (green open rhombuses) and $P = 20$ (red open circles), while (e) $\underline{L}_0$ and (f) $ZT$ as functions of the chemical potential ($\mu$) for finite periodic chains with (circles and up triangles) and without (squares and down triangles) Fano impurities at $k_BT = 0.01|t|$ (circles and squares) and $k_BT = 0.03|t|$ (up and down triangles). These chains are connected to two semi-infinite leads at their ends, as shown in the insets of (c) and (e).

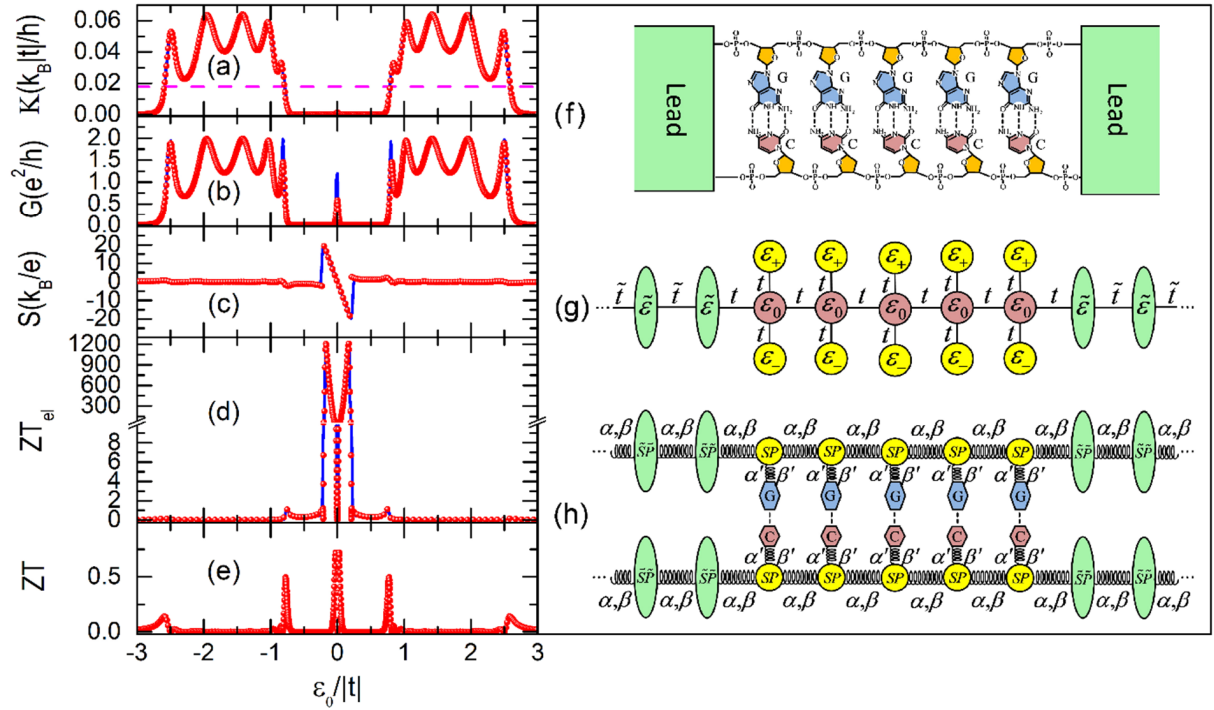

**FIG. 2**. (a) Thermal conductance ($K$) by electrons (red spheres) and by phonons (magenta dashed line), (b) electric conductance ($G$), (c) Seebeck coefficient ($S$), (d) electronic ($ZT_{el}$) and (e) full ($ZT$) thermoelectric figures of merit as functions of the nucleobase self-energy ($\varepsilon_0$) for a poly(G)-poly(C) DNA chain of five nucleotides connected to two semi-infinite leads as in Ref. [9], whose results are illustrated as blue lines in (a-d). Schematic representations of (f) a periodic poly(G)-poly(C) DNA double chain terminated by two leads, (g) fishbone model for electrons and (h) two-site coarse-grain model for phonons, both including semi-infinite periodic leads at their ends.

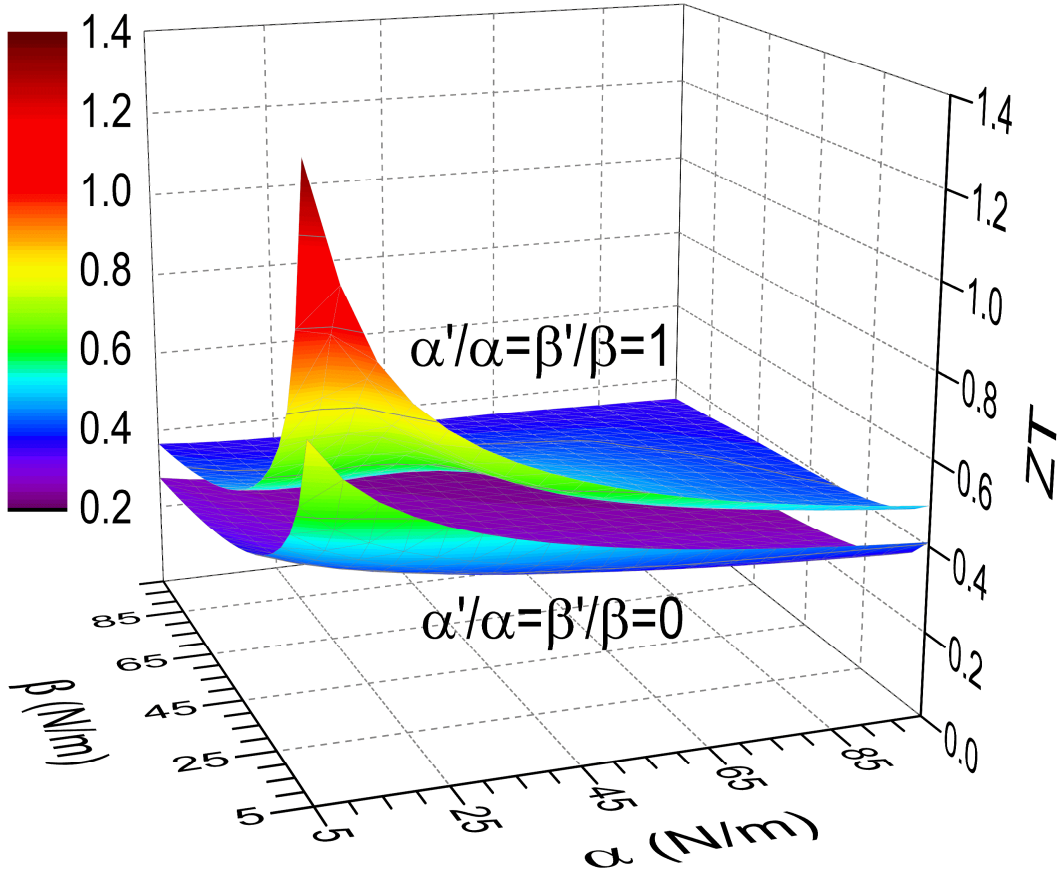

**FIG. 3**. Dimensionless thermoelectric figure of merit (*ZT*) as a function of the central ($\alpha$) and non-central ($\beta$) restoring force constants with $\varepsilon_0 = -0.027|t|$ for a poly(G)-poly(C) DNA chain of five GC base pairs connected to two semi-infinite atomic leads with the same parameters of FIG. 2.

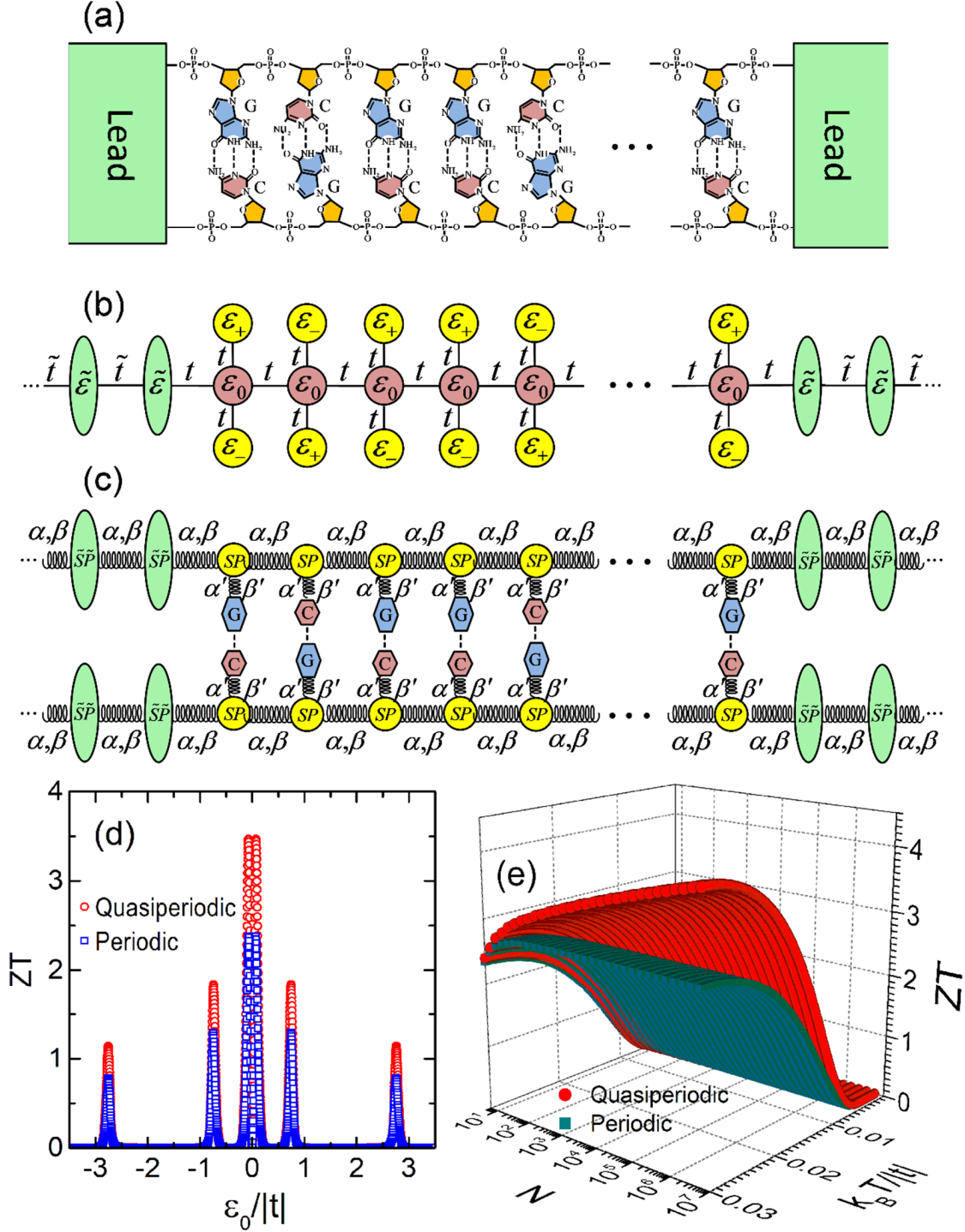

**FIG. 4**. Sketches of (a) a quasiperiodically ordered poly(G)-poly(C) DNA double chain terminated by two leads, (b) fishbone model for electrons and (c) two-site coarse-grain model for phonons, both including semi-infinite periodic leads at their ends. (d)The dimensionless thermoelectric figure-of-merit (ZT) versus the nucleobase self-energy ($\varepsilon_0$) normalized by the hoping parameter ($t$) for periodic (blue open squares) and quasiperiodic (red open circles) poly(G)-poly(C) double chains. (e) Variation of the thermoelectric figure-of-merit (ZT) as functions of the system length (N) and temperature (T) for both periodic (blue solid squares) and quasiperiodic (red solid circles) poly(G)-poly(C) double chains as (d).

16